\documentclass[twocolumn,prb,aps,showpacs,superscriptaddress]{revtex4-1}

\usepackage{siunitx}
\usepackage{physics}
\usepackage{nicefrac}
\usepackage{graphicx}
\usepackage{amsmath}
\usepackage{float}
\usepackage{xcolor} 
\usepackage[colorlinks, linkcolor=blue, citecolor=blue]{hyperref}

\begin{document}

\title{Magnetic domain walls in strain-patterned ultrathin films}

\author{Aurore Finco}

\author{Marco Perini}
\email{mperini@physnet.uni-hamburg.de}

\author{Andr\'{e} Kubetzka}

\author{Kirsten von Bergmann}

\author{Roland Wiesendanger}

\affiliation{Department of Physics, University of Hamburg, D-20355 Hamburg,
  Germany}

\date{\today}

\begin{abstract}

  We present a comparison of the characteristics of the magnetic domain walls in an atomic monolayer of Co on Pt(111) and a Ni/Fe atomic bilayer on Ir(111), based on spin-polarized scanning tunneling microscopy measurements. In both cases, the films exhibit a roughly triangular dislocation line pattern created by epitaxial strain relief, as well as out-of-plane ferromagnetic order. Domains with opposite magnetization are separated by domain walls with a unique rotational sense, demonstrating the important role of the Dzyaloshinskii-Moriya interaction induced by the Co/Pt and the Fe/Ir interfaces. The domain walls in Co/Pt(111) are straight and usually found in geometrical constrictions of the film, where they can minimize their length. In contrast, the domain walls in Ni/Fe/Ir(111) follow complicated paths, which can be correlated to the structural triangular pattern. The comparison between the two systems shows that the structural patterns, despite their similarity, have a different impact on the domain walls. In the Co/Pt(111) case, the magnetic state is not influenced by the dislocation line network, in contrast to the Ni/Fe/Ir(111) system in which the formation of the walls is favored at specific positions of the structural pattern.

\end{abstract}

\pacs{}

\maketitle

\section{Introduction}

Pinning of localized magnetic states, such as domain walls or skyrmions, is a common and important phenomenon in the study of magnetic structures at the nanoscale. It occurs mostly because of defects and inhomogeneities in the structure of magnetic films. In view of the development of spintronics devices like racetrack memories,\cite{parkin_magnetic_2008, fert_skyrmions_2013} it is crucial to reduce the pinning of the magnetic structures in order to reliably move them. For example, magnetic domain walls and skyrmions can be pinned at defects\cite{hanneken_pinning_2016, franken_domain-wall_2012} or at grain boundaries in alloys\cite{yu_pinning_1999} and sputtered films.\cite{woo_observation_2016, legrand_room-temperature_2017} As a consequence, larger current densities are necessary to counteract the pinning force and to move such magnetic objects, resulting in less energetically efficient devices.
On the other hand, the pinning of domain walls in artificially created notches is essential to guarantee a controlled motion of domain walls in ferromagnetic nanowires.\cite{hayashi_dependence_2006, atkinson_controlling_2008} Therefore, a deeper understanding of the pinning mechanisms for magnetic objects at the nanoscale is of utmost importance for novel spintronics devices.

In order to facilitate the identification of the mechanisms responsible for the pinning, we investigated the interaction of magnetic domain walls with well-defined structural features in epitaxial ultrathin films. The pinning of domain walls to strain-induced nanostructures was previously studied in FePt thin films on Pt(001),\cite{attane_domain_2001} where the structural changes are significant: up to \SI{3}{\nano\meter}-high steps appear. In the present work, we study out-of-plane ferromagnetic ultrathin films patterned by strain relief, the atomically thin Co monolayer on Pt(111) and the Ni/Fe atomic bilayer on Ir(111). In these systems, the observed pattern is created by lateral variations of the stacking of the atoms in the film. Compared to FePt/Pt(001), the structural changes of the film are much smaller but we show here that they can also have a large impact on the magnetic state.

We used spin-polarized scanning tunneling microscopy (SP-STM),\cite{wiesendanger_spin_2009} exploiting both its spatial resolution and its ability to measure non-collinear magnetic states. In previous studies, the detailed atomic arrangement in Co nanoislands and nanowires on Pt(111)\cite{lundgren_thin_2000, meier_spin-dependent_2006} and Ni/Fe nanoislands on Ir(111)\cite{iaia_structural_2016} has been determined. These measurements have also shown that both systems exhibit ferromagnetic order. However, the magnetic state of extended films has not been investigated so far in these systems and our work reveals in both cases the presence of domain walls with a unique rotational sense fixed by the large Dzyaloshiskii-Moriya interaction (DMI)\cite{dzyaloshinskii_1957, moriya_1960} induced at the Co/Pt\cite{pizzini_chirality-induced_2014, belmeguenai_interfacial_2015, yang_anatomy_2015, corredor_sempa_2017, simon_magnetism_2018} and the Fe/Ir\cite{heinze_spontaneous_2011} interfaces. The observed domain walls, although having comparable widths, show very different pinning properties in the two systems, shown by the different equilibrium wall positions.

\section{Experimental details}

Both the sample preparations and the measurements were performed in a multichamber ultrahigh vacuum system with a base pressure below \SI{e-10}{\milli\bar}. The cleaning of the substrates, the deposition of the metallic ultrathin films and the low-temperature STM measurements took place in separated chambers.

The Pt single crystal was cleaned by repeated cycles of Ar-ion sputtering at \SI{750}{\electronvolt} and annealing to temperatures of \SI{800}{\kelvin} for \SI{5} minutes. The Co was deposited at room temperature to minimize the intermixing with the substrate.\cite{lundgren_atomic-scale_1999} Typical deposition rates were around 0.07 atomic layers per minute.

The Ir substrate was prepared by cycles of Ar-ion sputtering at \SI{800}{\electronvolt} and annealing up to \SI{1500}{\kelvin} for \SI{90}{\second}. The Fe monolayer was deposited onto the Ir(111) surface about \SI{5}{min} after the annealing, which means that the substrate temperature was still elevated. This is necessary to achieve step-flow growth of a pseudomorphic fcc-stacked Fe monolayer.~\cite{heinze_spontaneous_2011} The typical deposition rate was around 0.2 atomic layers per minute. Once the sample had reached room temperature, the Ni layer was grown on top of the Fe layer at a slightly lower rate, around 0.15 atomic layers per minute.

The measurements were performed in a low-temperature STM with a base temperature of \SI{4}{\kelvin}, using a chemically etched Cr bulk tip. Superconducting coils allow to apply an external out-of-plane magnetic field up to \SI{9}{\tesla}. 

\section{Structure}

\begin{figure}[h!]
  \centering
  \includegraphics{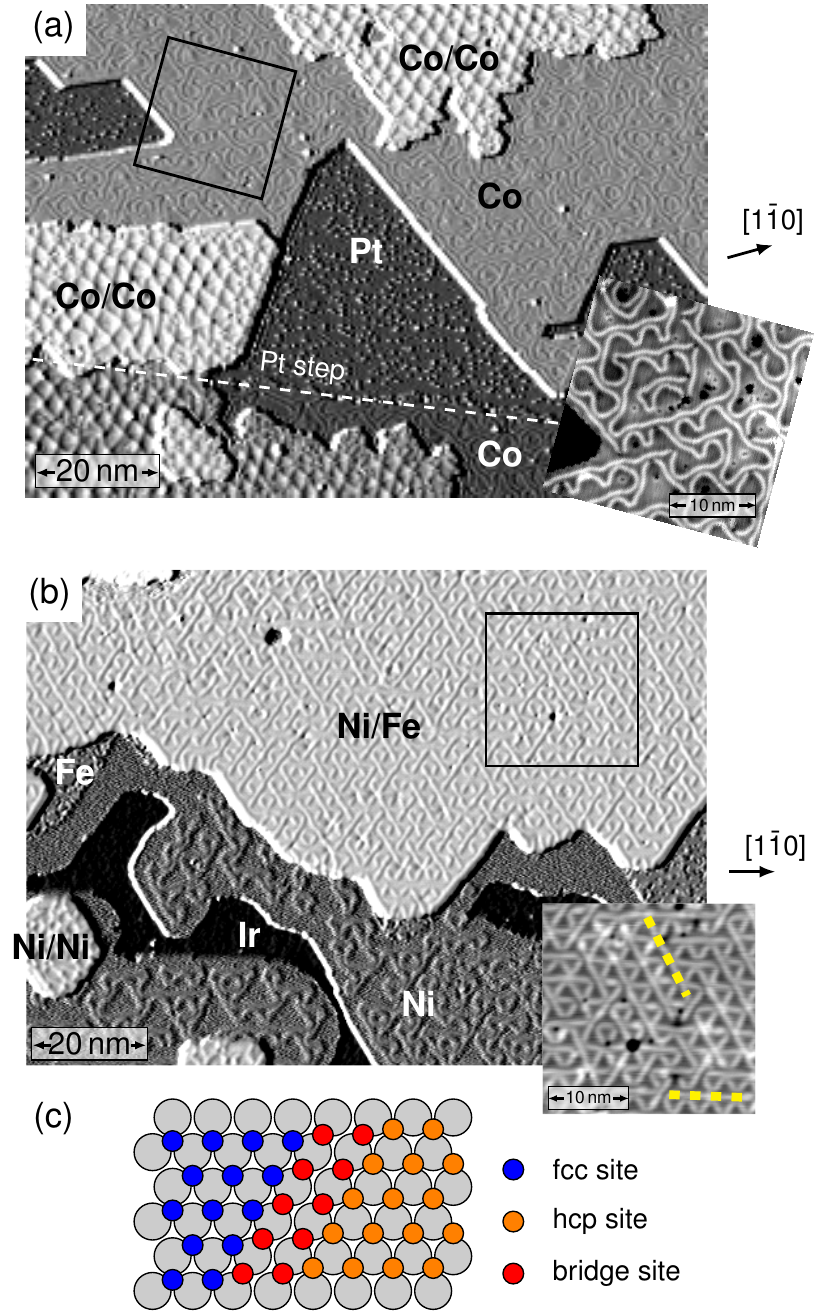}
  \caption{Structure of the ultrathin films. (a) STM constant-current map of a Co/Pt(111) sample, with an inset showing the details of the dislocation lines network. The Co coverage is about 1 atomic layer. (b) STM constant-current map of a Ni/Fe/Ir(111) sample. The Fe coverage is about 0.7 atomic layer and the Ni coverage about 0.9 atomic layer. The yellow lines in the inset mark long bridge lines. The inset shows in more details the dislocation lines. (c) Sketch showing the arrangement of the atoms corresponding to the structural patterns. The overview images were partially differentiated to improve the visibility of the topographic features. Measurement parameters: a main:~\SI{250}{\milli\volt}, a inset:~\SI{-1}{\volt}, b:~\SI{100}{\milli\volt}, all: \SI{1}{\nano\ampere}, \SI{4}{\kelvin}, Cr bulk tip.}
  \label{fig:structure}
\end{figure}

Figure~\ref{fig:structure} shows a comparison between the constant-current images of a Co on Pt(111) sample (Fig.~\ref{fig:structure}a) and a Ni/Fe on Ir(111) sample (Fig.~\ref{fig:structure}(b)).
\begin{figure}[h!]
  \centering
  \includegraphics{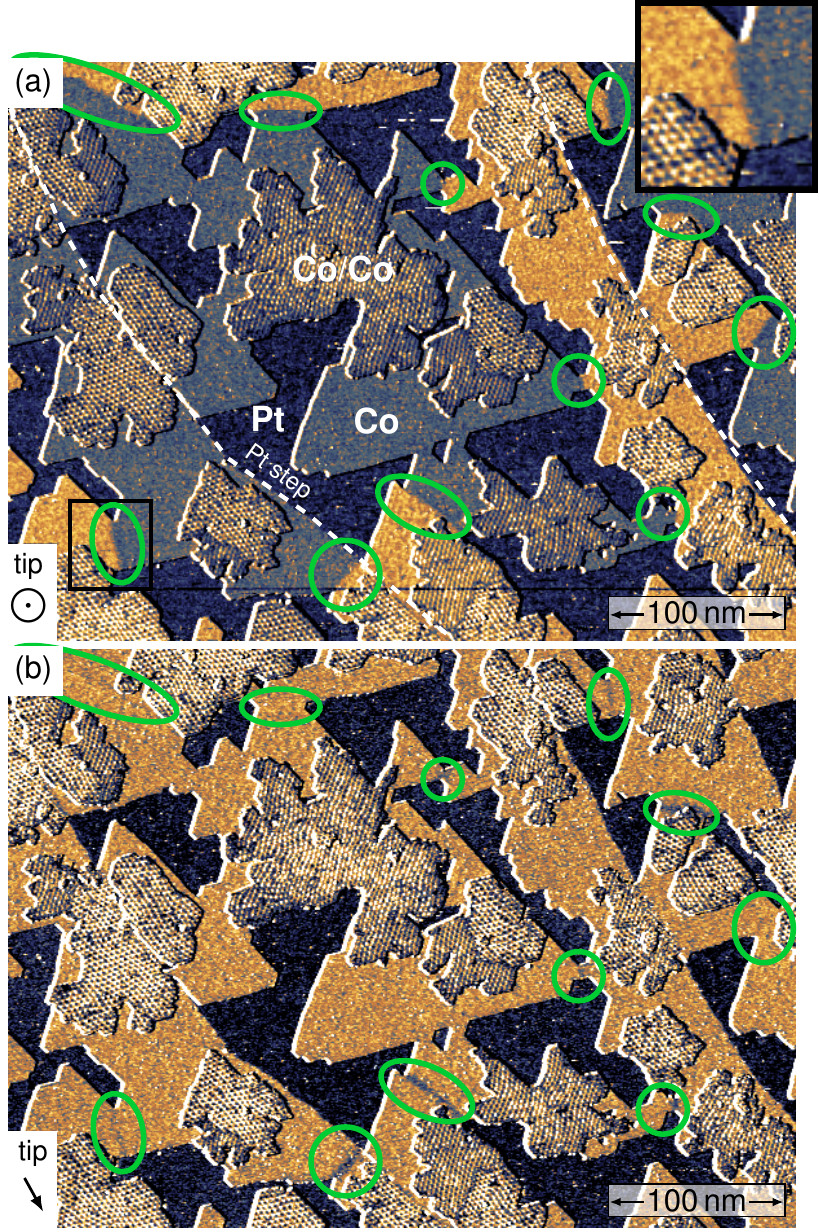}
  \caption{Differential conductance maps of a Co/Pt(111) sample, measured with a tip sensitive to the out-of-plane (a) and in-plane~(b) sample magnetization components, respectively. The domain walls, indicated by the green ellipses, are located at geometrical constrictions, a closer view of one wall is shown in the inset.
    Measurement parameters: \SI{250}{\milli\volt}, \SI{1}{\nano\ampere}, \SI{4}{\kelvin}, \SI{0}{\tesla}, Cr bulk tip.}
  \label{fig:oop_rot_sense_CoPt}
\end{figure}
 In both cases, the large lattice mismatch between the film and the substrate, of 9.4\% for the Co/Pt interface\cite{lundgren_thin_2000} and 8.2\% between Ni and Fe/Ir respectively, produces epitaxial strain which is relieved by the formation of various structural patterns. In the case of Co/Pt(111), the Co monolayer exhibits an irregular triangular pattern made of areas in which the Co atoms are located either in the fcc or the hcp hollow sites, separated by lines of atoms located in bridge sites.\cite{lundgren_thin_2000} These bridge lines appear brighter in the constant-current image of the inset of Fig.~\ref{fig:structure}(a). A sketch of this atom arrangement is shown in Fig.~\ref{fig:structure}(c). The fcc stacking is prefered for the Co atoms,\cite{lundgren_thin_2000} which means that the up-pointing triangles in the inset of Fig.~\ref{fig:structure}(a) are the fcc areas and the smaller down-pointing triangles the hcp areas. The width of the bridge lines is about \SI{0.7}{\nano\meter}. In the constant-current map of Fig.~\ref{fig:structure}(a) also some Co double layer areas are visible, which display a hexagonal structure corresponding to a Moir\'{e} pattern.\cite{lundgren_thin_2000}

Figure~\ref{fig:structure}(b) shows a constant-current image of Ni/Fe/Ir(111). The Fe monolayer on Ir(111) grows pseudomorphically in fcc stacking,\cite{heinze_spontaneous_2011} whereas an irregular triangular dislocation line pattern is present in the Ni layer deposited on top (and to a lesser extent in the Ni monolayer on Ir(111)).\cite{iaia_structural_2016} The atom arrangement producing this triangular structure is analogous to the Co/Pt(111) system (see Fig.~\ref{fig:structure}(c)) and similarly, the bridge lines appear bright in the constant-current image of the inset of Fig.~\ref{fig:structure}(b), with a width of again about \SI{0.7}{\nano\meter}. However, from the comparable size of the up- and down-pointing triangles in the structure, it is deduced that the fcc and the hcp stackings are roughly energetically equivalent in this case\cite{iaia_structural_2016} (see inset in Fig~\ref{fig:structure}(b)). The triangular structure is not fully regular. Some of the bridge lines are longer and adjoin several fcc or hcp areas. Two of these lines are marked with a yellow dashed line in the inset in Figure~\ref{fig:structure}(b). These long bridge lines play a crucial role in the pinning of the domain walls in Ni/Fe/Ir(111), as discussed in the next sections.

\section{Magnetism}

Previous SP-STM investigations, mainly focussed on nanowires and nanoislands, have demonstrated that both the atomic Co monolayer on Pt(111) and the Ni/Fe atomic bilayer on Ir(111) are ferromagnetic with out-of-plane anisotropy.\cite{meier_spin-dependent_2006, iaia_structural_2016}

\begin{figure}[h!]
  \centering
  \includegraphics{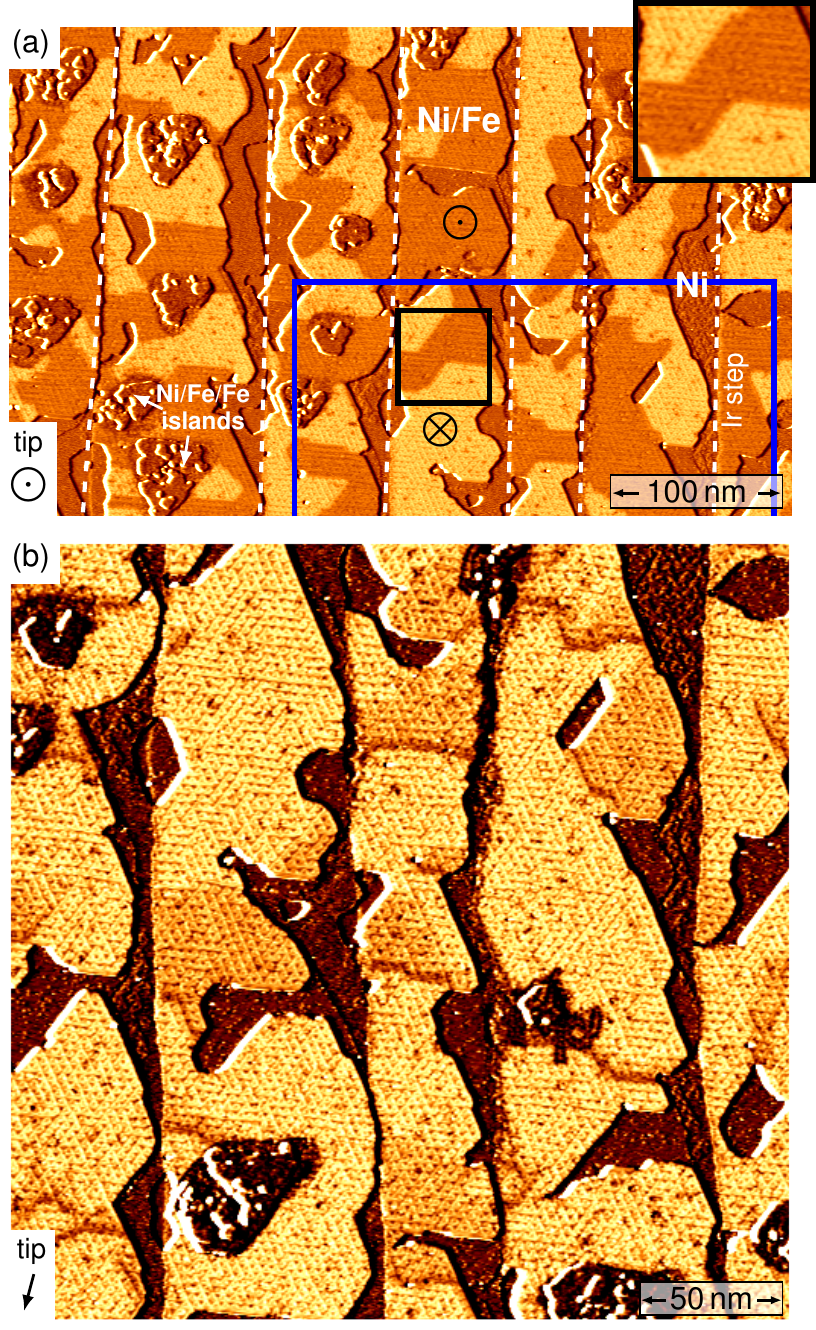}
  \caption{Spin-resolved differential conductance maps of a Ni/Fe/Ir(111) sample measured with an out-of-plane sensitive tip (a) and a tip sensitive to both the out-of-plane and an in-plane component of the magnetization ((b), the in-plane component is indicated with the arrow). The blue box in (a) marks part of the area shown in (b). Two domain walls are shown in a closer view in the inset.
    Measurement parameters: \SI{100}{\milli\volt}, \SI{1}{\nano\ampere}, \SI{4}{\kelvin}, a:~\SI{0}{\tesla}, b:~\SI{200}{\milli\tesla}, Cr bulk tip.}
  \label{fig:oop_rot_sense_NiFe}
\end{figure}

Figure~\ref{fig:oop_rot_sense_CoPt}(a) shows a spin-resolved differential conductance map of a sample with Co monolayer and Co double layer areas on Pt(111), measured with an out-of-plane sensitive magnetic tip. Large oppositely magnetized domains are visible, separated by domain walls in the Co monolayer. A closer view of the wall marked with the black square is shown in the inset. The data shows that the domain walls are preferentially located in geometrical constrictions of the Co film. This 
results from a positive energy cost per unit length of a domain wall, 
and the position of the walls in constrictions allows to minimize their length and thus their energy.

In order to determine if the domain walls have a unique rotational sense and thus know if the DMI plays a role in this system, knowledge about the in-plane components of the walls is required as well. We have modified the tip apex outside the image area by gentle indentations with the sample until a magnetic in-plane sensitivity was achieved.
The same area of Fig.~\ref{fig:oop_rot_sense_CoPt}(a) is now imaged with an in-plane tip and shown in Fig.~\ref{fig:oop_rot_sense_CoPt}(b). The out-of-plane domains cannot be discriminated anymore, but the in-plane components of the domain walls along the in-plane magnetized tip are visible.
Note that the observed domain wall contrast in this image does not only originate from the tunneling magnetoresistance (TMR)\cite{wiesendanger_spin_2009} but also from an additional electronic contribution\cite{bode_magnetization-direction-dependent_2002, hanneken_electrical_2015} (see the Supplemental Material for more information). The correlation between Figs.~\ref{fig:oop_rot_sense_CoPt}(a) and (b) allows to conclude that the domain walls have a unique rotational sense, more details can be found in the Supplemental Material. This shows that the DMI at the Co/Pt interface is significant, in agreement with previous work.\cite{pizzini_chirality-induced_2014, belmeguenai_interfacial_2015, yang_anatomy_2015, corredor_sempa_2017, simon_magnetism_2018} Because such an interfacial-DMI stabilizes N\'{e}el walls with fixed rotational sense\cite{heide_dzyaloshinskii-moriya_2008, thiaville_dynamics_2012} over Bloch walls, we can conclude that the magnetization in the wall is cycloidal. From this, we can deduce the tip magnetization axis during the measurement, as indicated by the arrow of Fig.~\ref{fig:oop_rot_sense_CoPt}(b).

Figure~\ref{fig:oop_rot_sense_NiFe} shows a similar experiment for Ni/Fe/Ir(111). The differential conductance map (a) is measured with an out-of-plane sensitive tip and reveals the presence of out-of-plane oppositely magnetized domains. However, in contrast to the case of the Co monolayer on Ir(111), the domain walls do not minimize their lengths in geometrical constrictions of the Ni/Fe film but instead follow more complicated paths. A closer view of two walls (marked with the black box) is shown in the inset and suggests that the unusual paths followed by these walls are related to the structural strain-relief-induced pattern. We will look at this more closely in the next section.\\
To investigate the characteristics of the domain walls, we also measured the in-plane components of the magnetization in the domain walls in the Ni/Fe/Ir(111) system. Figure~\ref{fig:oop_rot_sense_NiFe}(b) is a differential conductance map of an area overlapping with the one shown in (a) (see the blue box). In this case, the magnetization at the tip apex is canted and thus the tip is sensitive to both the out-of-plane component and an in-plane component of the magnetic state of the sample (see Supplemental Material). The oppositely magnetized out-of-plane domains are still visible, they appear bright or slightly dark. Note that an out-of-plane magnetic field was applied to the sample between the measurements (a) and (b), which induced the movement of some of the walls. \\
The tip sensitivity also allows to observe the in-plane component of the magnetization in the domain walls, which manifests itself as a brighter or darker contrast at the position of the walls. However, an additional electronic contribution is also present as an offset, similar to the case of Co/Pt(111) (see Supplemental Material), which makes the domain walls appear either very dark or hardly visible. Since the order bright domain/very dark wall/slightly dark domain/hardly visible wall is repeated everywhere without changing the wall orientation (mostly perpendicular to the Ni stripe), we can conclude that the magnetic domain walls have a unique rotational sense in Ni/Fe/Ir(111). Thus also in this system, the DMI is involved in their stabilization, which is not surprising because of the large DMI induced by the Fe/Ir(111) interface.\cite{heinze_spontaneous_2011} 

\section{Characteristics of the domain walls}

\begin{figure}[h!]
  \centering
  \includegraphics{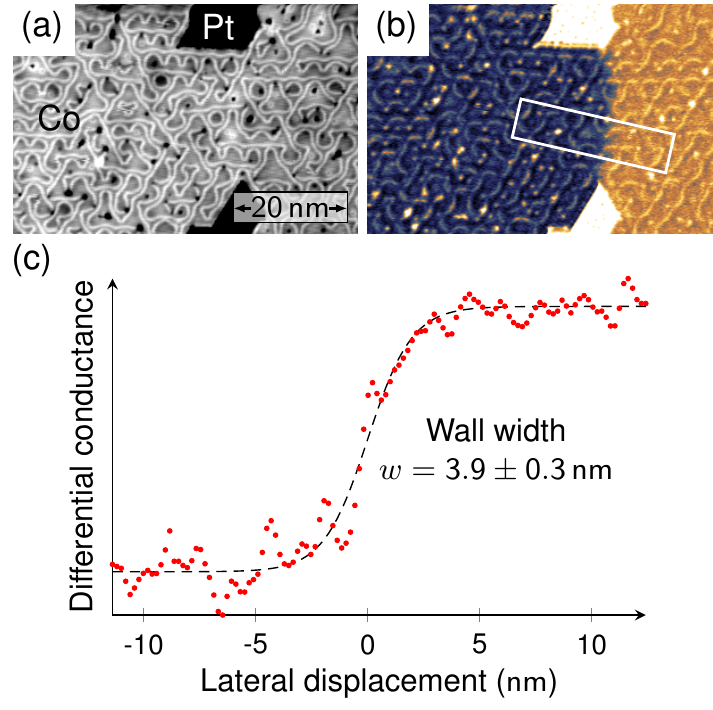}
  \caption{STM constant-current map~(a) and the simultaneously recorded spin-polarized differential conductance map~(b) of a Co monolayer on Pt(111). The signal in the white box is averaged in the short direction and plotted against the long direction in~(c). The black dashed line is a fit to the data points using Eq.~(\ref{eq:profile}) for a fully out-of-plane sensitive magnetic tip.
    Measurement parameters: \SI{-770}{\milli\volt}, \SI{1}{\nano\ampere}, \SI{4}{\kelvin}, \SI{0}{\tesla}, Cr bulk tip.}
  \label{fig:DWprofile_CoPt}
\end{figure}

The profile of a domain wall in a magnetic ultrathin film with uniaxial anisotropy can be computed by minimizing the following micromagnetic energy functional:\cite{heide_dzyaloshinskii-moriya_2008} 
\begin{equation}
  \label{eq:energy_func}
  E[\varphi] = \int_{-\infty}^{+\infty} \left[A\left(\dv{\varphi}{x}\right)^2 + D \dv{\varphi}{x} + K \sin^2 \varphi \right] \dd{x}
\end{equation}
where $\varphi(x)$ is the angle between the normal to the surface and the magnetization at the coordinate $x$, $A$ is the exchange stiffness, $D$ is the DMI constant and $K$ the effective uniaxial anisotropy. In order to keep the model simple, the magnetic parameters are assumed to be spatially uniform, only averaged values are considered. This is actually an approximation in Co/Pt(111) and in Ni/Fe/Ir(111) because of the strain-relief-induced pattern. Indeed, since this pattern corresponds to variations of the lateral positions of the atoms, it is expected that a spatial modulation of the magnetic parameters in the film is induced.\\
We impose the presence of two oppositely magnetized out-of-plane domains by choosing the boundary conditions for the angle $\varphi$:
\begin{equation}
  \label{eq:lim}
  \lim_{x\to+\infty}\varphi(x) = 0, \lim_{x\to-\infty}\varphi(x) = \pi  
\end{equation}
The minimization of the energy\cite{heide_dzyaloshinskii-moriya_2008} leads to this expression for the profile of the domain wall:
\begin{equation}
  \label{eq:profile}
  \cos \varphi(x) = \pm \left(\tanh\left(\frac{x}{\nicefrac{w}{2}}\right)\right)
\end{equation}
with $w = 2\sqrt{\nicefrac{A}{K}}$. The DMI does not influence the shape of the wall, but it changes its energy per unit length:
\begin{equation}
  \label{eq:energy_dw}
  E = 4\sqrt{AK} \pm \pi D
\end{equation}

\begin{figure}[h]
  \centering
  \includegraphics{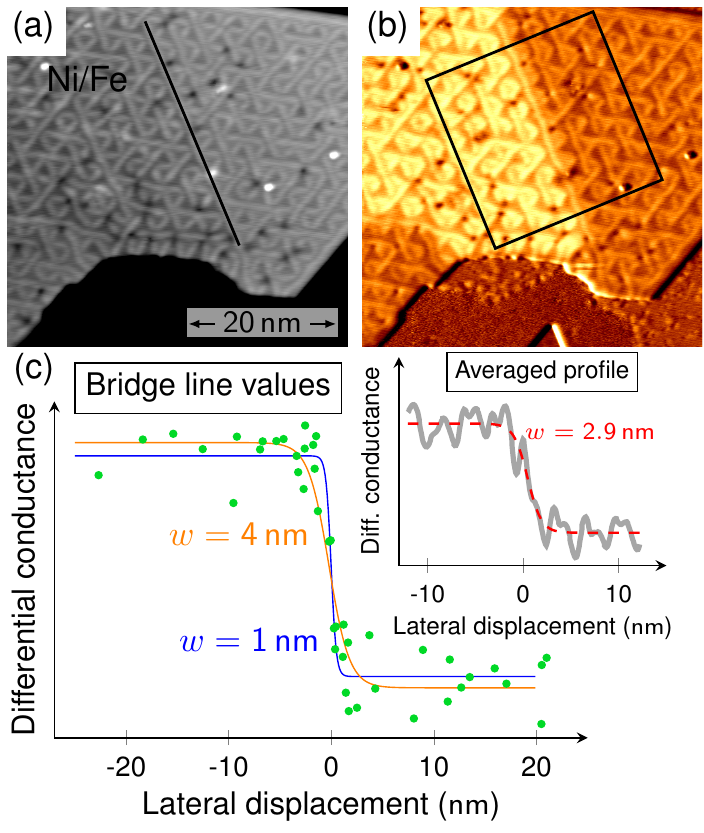}
  \caption{STM constant-current map~(a) and the simultaneously recorded spin-polarized differential conductance map~(b) of a Ni/Fe film on Ir(111). The plot~(c) shows the values of the differential conductance in image~(b) at points located on the bridge lines as a function of the distance to the long bridge line marked in image (a). Two domain wall profiles were fitted to the data (green dots) with the constraint that the wall width is either 1 or \SI{4}{\nano\meter} (blue and orange lines respectively). In the inset, the differential conductance profile (averaged over the area in the box indicated in image~(b)) is shown, as well as a fit (red line). The fit gives a wall width of \SI{2.9}{\nano\meter}.
    Measurement parameters: \SI{100}{\milli\volt}, \SI{1}{\nano\ampere}, \SI{4}{\kelvin}, \SI{0}{\tesla}, Cr bulk tip.}
  \label{fig:DW_position_NiFe}
\end{figure}

\begin{figure*}
  \centering
  \includegraphics{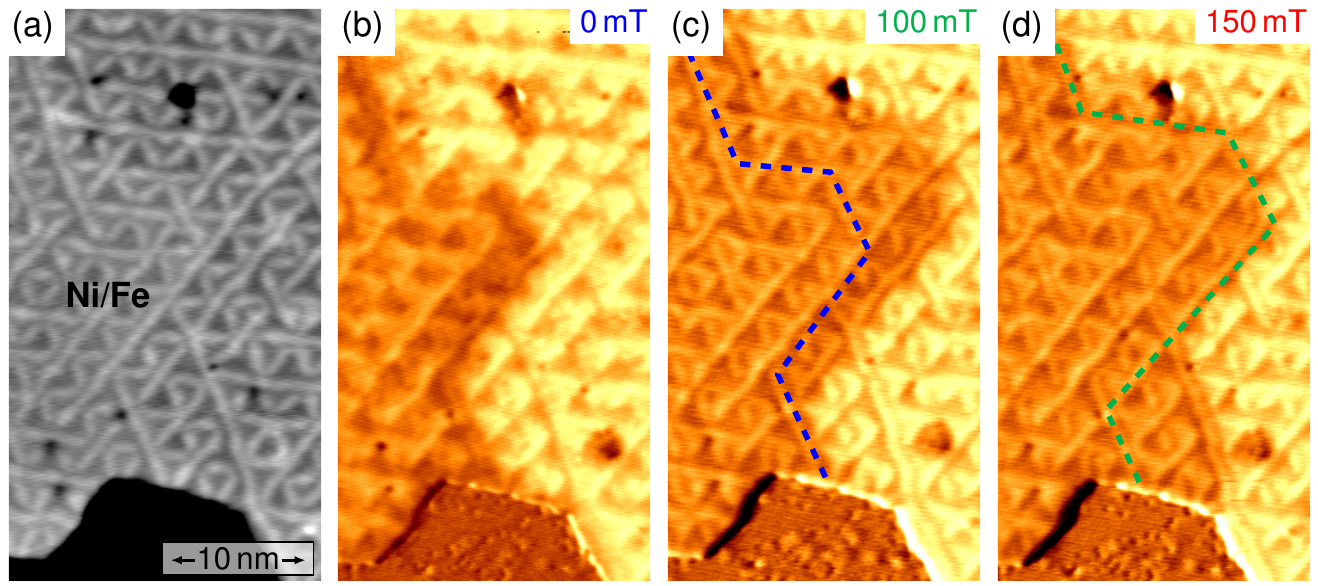}
  \caption{Constant-current map~(a) and spin-resolved differential conductance maps~(b)-(d) showing the field dependence of the positioning of a domain wall in Ni/Fe/Ir(111). The constant-current map shows the details of the dislocation line pattern. When the field increases, the domain on the left grows and the domain wall jumps from one bridge line to the next. The tip is sensitive to the out-of-plane component of the sample magnetization.
    Measurement parameters: \SI{100}{\milli\volt}, \SI{1}{\nano\ampere}, \SI{4}{\kelvin}, Cr bulk tip.}
  \label{fig:field_dep_NiFe}
\end{figure*}

A detailed analysis of a domain wall in the Co monolayer on Pt(111) is shown in Figure~\ref{fig:DWprofile_CoPt}. This wall is located in a constriction which is about \SI{20}{\nano\meter} wide. 
The profile in (c) is extracted from the differential conductance map measured at \SI{-770}{\milli\volt}, a bias voltage at which the signal change due to the structural pattern is small compared to that of magnetic origin. The data confirms the wall width of about \SI{4}{\nano\meter} found previously in  nanowires of Co on Pt(111).\cite{meier_spin-dependent_2006} The comparison between the STM constant-current map in Fig.~\ref{fig:DWprofile_CoPt}(a) and the differential conductance map in Fig.~\ref{fig:DWprofile_CoPt}(b) shows that the position of the domain wall is determined by the geometry of the local constriction in the Co monolayer film rather than by the structural pattern.

A close view of a domain wall in the Ni/Fe bilayer on Ir(111) is presented in Fig.~\ref{fig:DW_position_NiFe}. We find that is does not take the shortest path across the stripe, but instead it runs at an angle of about 30 degrees with respect to the stripe. The position of the domain wall seems to correlate with the position of a particularly long straight bridge line, suggesting that the magnetic state pins to the structural pattern. To investigate this further, we need to determine the center of the domain wall with respect to the bridge line. 
In this case, the structural triangular pattern is clearly visible at any bias voltage in the differential conductance, with a signal change on the same order of magnitude as the magnetic contrast. Therefore, even after averaging in the direction parallel to the wall (over the area indicated in~\ref{fig:DW_position_NiFe}(b), like it is done in Fig.~\ref{fig:DWprofile_CoPt}), the structural pattern produces a strong modulation of the differential conductance, as shown by the gray line in the inset of Fig.~\ref{fig:DW_position_NiFe}(c). Fitting Eq.~(\ref{eq:profile}) for a magnetic sensitivity of the tip fixed out-of-plane to such a profile can give a rough estimate of the domain wall width. However, the obtained position of the wall center is not reliable because of the distortions induced by the structure. We gain additional information with a different procedure. We reduce the modulation of the differential conductance signal, which is produced by the different stackings of the atoms, by selecting data points belonging to only one specific stacking, i.e. either to fcc, hcp or to bridge lines. We chose points belonging to bridge lines, and then plot the corresponding differential conductance as a function of the distance from the long bridge line indicated in \ref{fig:DW_position_NiFe}(a). The result is shown in Fig.~\ref{fig:DW_position_NiFe}(c). 
\\
The obtained profile is not completely smooth and does not allow to determine precisely the wall width. The shape from Eq.~(\ref{eq:profile}) was nevertheless fitted to the data, with the constraint that the tip is out-of-plane sensitive and for a wall width of either \SI{1}{\nano\meter} (blue line) or \SI{4}{\nano\meter} (orange line), but no constraint on the position of the wall. The results show that the center of the wall, which is the area where the magnetic moments are pointing in-plane, is located at 0, i.e. at the position of the bridge line, for both fitted profiles. The wall width appears to be in the range between \SI{1}{\nano\meter} and \SI{4}{\nano\meter}, which is in agreement with the value of \SI{2.9}{\nano\meter} obtained on the averaged profile presented in the inset. We conclude that indeed for Ni/Fe/Ir(111) the bridge sites act as pinning lines for the domain walls.\\
The experiment presented in Fig.~\ref{fig:field_dep_NiFe} illustrates the strong pinning of the domain walls to the bridge lines in the Ni/Fe bilayer on Ir(111). The STM constant-current map~(a) shows the structural pattern in an area exhibiting a wall with a complicated shape. The differential conductance maps~(b) to~(d) show the position of the domain wall when an external out-of-plane magnetic field is increased step by step. In the absence of magnetic field (Fig.~\ref{fig:field_dep_NiFe}(b)) the domain wall follows a path dictated by the bridge lines. When the external magnetic field is increased (Figs.~\ref{fig:field_dep_NiFe}(c) and (d)), the dark domain, which corresponds to the magnetization parallel to the field, grows. 
Thermal effects and the influence of the STM tip might also contribute to the observed domain wall motion. The wall successively jumps to the next long bridge lines and stays pinned. In addition to this experiment, the equilibrium positions of a significant number of domain walls were observed and it appeared that most of the walls in Ni/Fe/Ir(111) are pinned to a least one long bridge line. 

\section{Discussion}

In the two systems studied, the triangular pattern created by strain relief corresponds to spatial variations of the stacking of the atoms. It is known from previous studies that the stacking can have an influence on the magnetic state in ultrathin films.\cite{von_bergmann_influence_2015, gao_revealing_2008, romming_competition_2018} One can thus expect that the presence of a structural pattern in the film can induce a spatial modulation of the exchange stiffness, the DMI or the magnetic anisotropy, which determine the energy of the domain walls. From the expression of the energy given in Eq.~(\ref{eq:energy_dw}), the formation of a wall becomes more favorable if the exchange stiffness or out-of-plane anisotropy decreases or if the DMI constant increases. However, Co/Pt(111) and Ni/Fe/Ir(111) are very different materials. The stacking variation can modify the magnetic parameters differently in the two cases and even a similar change of one parameter can have a very different effect, since the domain walls are stabilized by the competition between several energy terms. Unfortunately, our STM experiments do not allow to determine which magnetic energy term among the exchange coupling, the DMI and the effective anisotropy is involved in the dominant pinning mechanism. Nevertheless, the presence of much longer walls in Ni/Fe/Ir(111) compared to Co/Pt(111) suggests that their formation might be more favorable in this system. 

In Pt/Co/Pt stripes irradiated with Ga,\cite{franken_domain-wall_2012} the pinning of domain walls was attributed to variations of the uniaxial anisotropy. The size of the irradiated region also plays an important role for the pinning. More generally, the pinning is stronger when the typical length of the structural variations is close to the size of the magnetic objects, as observed in the case of skyrmions pinned at the boundaries of grains with varying DMI.\cite{legrand_room-temperature_2017} In the Co/Pt(111) system, the sides of the approximately triangular fcc grains have an average size of about \SI{3}{\nm}. The hcp grains are smaller, with a size around \SI{2.5}{\nm}. In Ni/Fe/Ir(111), the sides of the triangles are roughly \SI{3.2}{\nm} long. In addition, the width of the bridge lines is the same in the two cases, \SI{0.7}{\nano\meter}. The grain size is thus rather close to the wall width of \SI{4}{\nano\meter} in Co/Pt(111), whereas in Ni/Fe/Ir(111), the wall width between \SI{1}{\nano\meter} and \SI{4}{\nano\meter} could be close to the width of the lines or the size of the grains. This effect of the grain size or the line width might contribute to the stronger pinning in Ni/Fe/Ir(111) than in Co/Pt(111).

Another geometrical effect might play a role in the pinning of the walls. The paths of the bridge lines are rather curved in Co/Pt(111) whereas the lines are very straight in Ni/Fe/Ir(111). It might be less favorable for the domain walls to follow a curved path than a straight one, which would also increase the pinning in Ni/Fe/Ir(111) compared to Co/Pt(111).

\section{Conclusion}

Our work shows that domain walls in epitaxial ultrathin films patterned by strain relief can have a very different behavior despite the similarity of the atom arrangement in the film. The Co monolayer on Pt(111) and the Ni/Fe bilayer on Ir(111) both exhibit an irregular triangular structural pattern which consists of alternating fcc and hcp areas separated by bridge lines. However, whereas the pattern does not affect the domain walls in Co/Pt(111), they are strongly pinned to the bridge lines in Ni/Fe/Ir(111). This pinning effect likely originates from a spatial modulation of the magnetic parameters (exchange coupling, DMI, effective anisotropy) induced by the stacking variations in the film. 
Their effects on the different magnetic energy contributions cannot be disentangled in our experiments. Nevertheless, the comparison between the two systems shows that the pinning is effectively smaller in Co/Pt(111) compared to Ni/Fe/Ir(111). Our work highlights how the choice of the appropriate materials, with specific pinning properties, is a crucial step towards the realization of novel spintronic devices.

\begin{acknowledgments}

  Financial support by the European Union via the Horizon 2020 research and innovation programme under grant agreement No. 665095 and by the Deutsche Forschungsgemeinschaft via SFB668-A8 is gratefully acknowledged.

\end{acknowledgments}

\bibliography{CoPt_NiFe}

\end{document}